\documentclass{JINST}
\let\ifpdf\relax
\pdfoutput=1
\usepackage{textcomp}
\newcommand{\Xray}{X\hbox{-}ray }
\newcommand{\Unit}[1]{\mathrm{\,#1}}
\title{Comparison of different sources for laboratory X-ray microscopy}

\author{Thomas Ebensperger$^{a,b}$\thanks{Corresponding author.}~, Philipp Stahlhut$^{a}$, Frank Nachtrab$^{b,c}$, Simon Zabler$^a$ and Randolf Hanke$^{a,b,c}$\\ 
\llap{$^a$}Chair for X-ray Microscopy, Julius-Maximilians-Universit\"at W\"urzburg\\
  Josef-Martin-Weg 63, 97074 W\"urzburg, Germany\\
\llap{$^b$}Fraunhofer Development Center X-Ray Technology (EZRT),\\
  Dr.-Mack-Str. 81, 90762 F\"urth, Germany\\
\llap{$^c$}Engineering of Advanced Materials, Friedrich-Alexander-Universit\"at Erlangen-N\"urnberg,\\
  Dr.-Mack-Str. 81, 90762 F\"urth, Germany\\ \\
  E-mail: \email{thomas.ebensperger@iis.fraunhofer.de}}

\abstract{This paper describes the setup of two different solutions for laboratory \Xray \hbox{microscopy} working with geometric magnification. One setup uses thin-film transmission targets with an optimized tungsten-layer thickness and the electron gun and optics of an electron probe \hbox{micro} analyzer to generate a very small \Xray source. The other setup is based on a scanning electron microscope and uses microstructured reflection targets. We also describe the structuring process for these targets. \par
In both cases we show that resolutions of $100\Unit{nm}$ can be achieved. Also the possibilities of computed tomography for 3D imaging are explored and we show first imaging examples of high-absorption as well as low-absorption specimens to demonstrate the capabilities of the setups.}

\keywords{X-ray microscopy, nano-focus X-ray source, laboratory setup}

\begin{document}
\section{Introduction}
Over the past decades, the spatial resolution in \Xray computed tomography (CT) systems set up at synchrotron beamlines reached $50\Unit{nm}$ in lensless diffraction imaging \cite{bib1} and $30\Unit{nm}$ using \Xray optics like Fresnel zone-plates \cite{bib2}. Similar reports from laboratory setups are exceptionally rare. In principle, optics (e.g.\ zone lenses) can be used in combination with high-power sources to realize specifications comparable to synchrotron beamlines (although with lower photon flux) \cite{bib3} but this solution comes with several constraints, mainly a.\ \Xray optics are expensive and very sensitive to vibration, b.\ such setups are designed for a narrow band of \Xray wavelengths, whereby the available photon energies generally remain below $10\Unit{keV}$. A possibility to mend these constraints and to perform imaging at higher energies is working in geometric magnification as it is generally the case for typical laboratory micro CT systems. A physical limit for the resolution of such a system is the size of the \Xray source spot, which is around 1\,\textmu m. Here, we present two laboratory-based \Xray setups with small source sizes and resolution below $100\Unit{nm}$ and compare their performance to synchrotron imaging and other \Xray microscopy techniques.

\section{Experimental Setups}
\begin{figure}
 \centering
 \includegraphics[width=0.9\columnwidth]{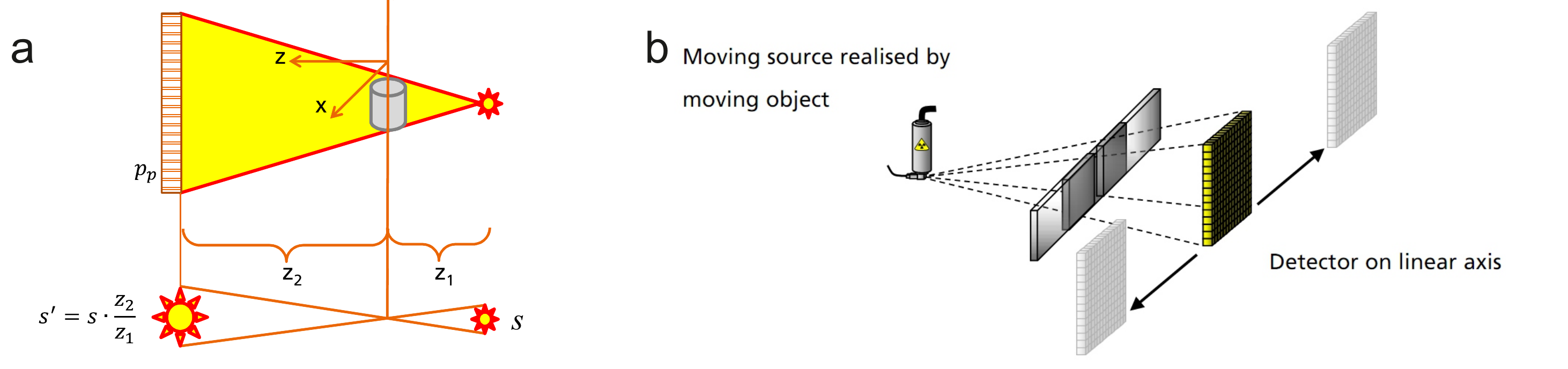}
 \caption{\textbf{a)} The principle of geometric magnification. Both the effective pixel size and the projection of the \Xray source size onto the imaging plane limit the resolution. \textbf{b)} The principle of the used laminographic imaging mode. Different views are captured by shifting object and detector and keeping the position of the source constant.}
 \label{fig:GeoLami}%
\end{figure}

When using the geometric magnification mode (see a schematic in fig.\ \ref{fig:GeoLami}a) and a pixelated \Xray detector for microscopy, there are two main resolution constraints: the finite sampling due to the finite pixel size and the projection of the \Xray source spot onto the imaging plane. When working with very high magnification the ultimate resolution limit of the presented systems is the size of the \Xray focal spot. The generation of such small \Xray source sizes can be achieved using a highly focused electron beam hitting a specially designed metal target. The materials chosen in the presented setups are tungsten for its high bremsstrahlung generation due to high atomic number and molybdenum for its characteristic lines that can be sufficiently excited with a $30\Unit{keV}$ electron beam.
\subsection{Transmission Setup}
The \Xray microscope set up at the Fraunhofer Development Center \Xray Technology in F\"urth is based on the electron gun and optics of an electron probe micro analyzer \cite{bib4}, where the sample chamber has been replaced by a transmission target. The thickness of the \Xray source layer (tungsten) has been optimized for high resolution using Monte Carlo simulations \cite{bib5}. Due to the low photon flux, a photon counting detector is required. We use the Quad version of the Medipix2~MXR detector with $512 \times 512$ pixels (pitch 55\,\textmu m) and a 300\,\textmu m silicon sensor layer. Magnifications up to $M = 1000$ can be achieved. Due to technicalities the use of a rotational axis for 3D imaging is not possible in this setup. So to realize 3D imaging, a laminography setup (see fig.\ \ref{fig:GeoLami}b) was chosen. Different views are realized by shifting detector and object on linear axes. This corresponds to a computed tomography setup with limited rotation angle (see also \cite{bib6}). The laminography angle was $\pm3.3^\circ$ at the chosen magnification.

\subsection{Reflection Setup}
\begin{figure}
		\centering
		\includegraphics[width=0.8\textwidth]{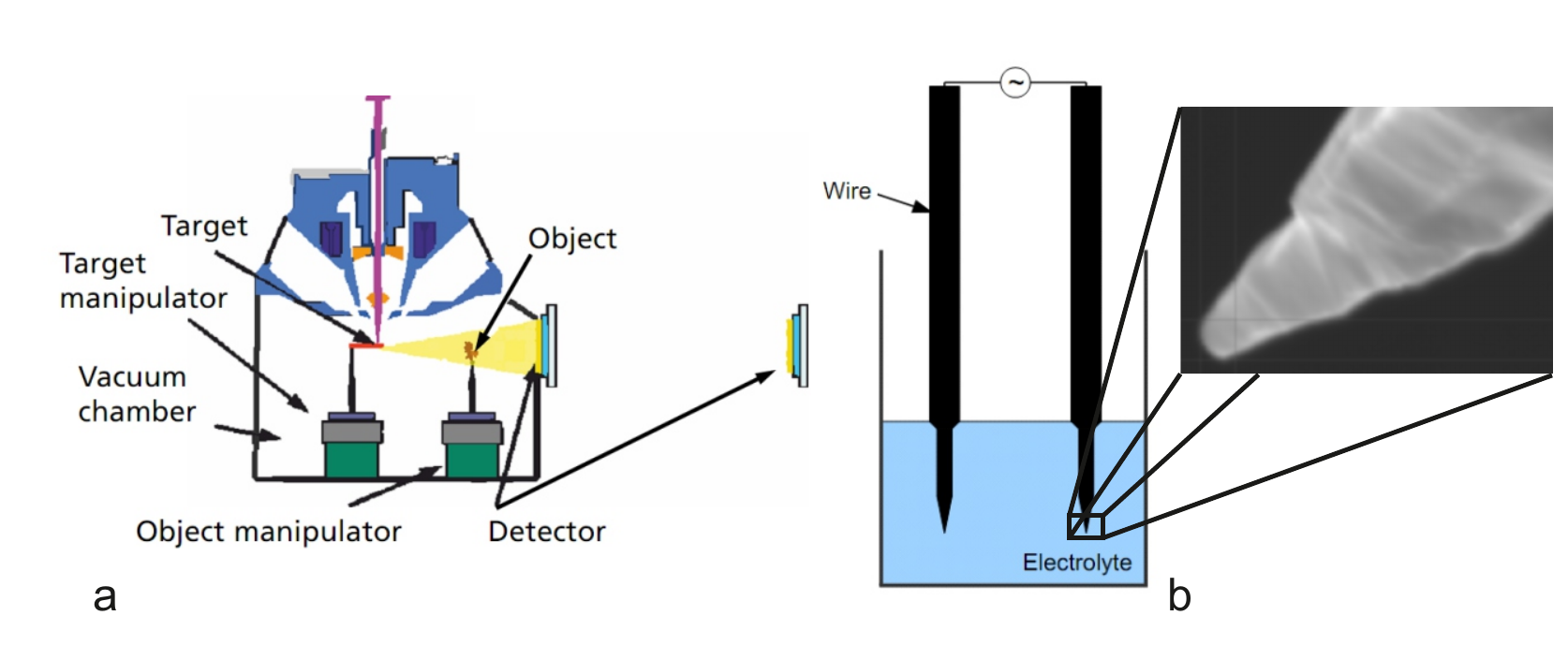}
		\caption{ \textbf{a)} Schematic of the reflection setup. Both target and object are inside a vacuum chamber. The detector is outside. \textbf{b)} Schematic of the electrochemical microstructuring process for manufacturing the reflection targets. Metal wires of molybdenum and tungsten are etched in NaOH  \cite{bib7}. The tip shown in the SEM image has a tip radius below 70\,nm.}
		\label{fig:Ref}
\end{figure}

To realize a setup with a rotational axis to have the possibility to perform a full CT scan, a horizontal primary beam offers far more convenience than a system with a vertical primary beam. We modified a JEOL JSM-7100F to serve as an \Xray microscope at W\"urzburg University. The sample stage has been replaced by a customized stage for both reflection target and object (see fig.\ \ref{fig:Ref}a). The \Xray detector is placed outside the vacuum chamber and is a $2\times3$ Medipix2 MXR with a $1\Unit{mm}$ cadmium telluride sensor layer. The key to the generation of small \Xray source sizes in a reflection target is---in addition to a small electron focus---the reduction of the physical extent of the \Xray generation volume. To achieve this, very sharp metal needles with a tip curvature below $70\Unit{nm}$ are used (see fig.\ \ref{fig:Ref}b).
 
\section{Results}
\begin{figure}
 \centering
 \includegraphics[width=\columnwidth]{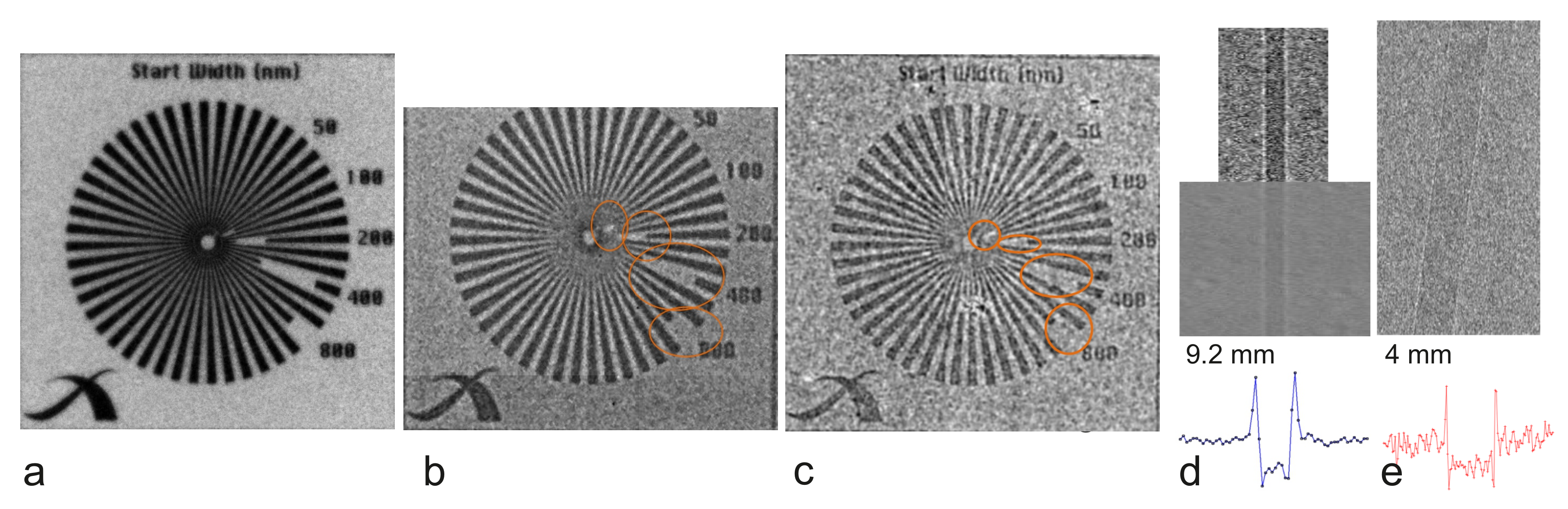}
 \caption{A test pattern imaged with the Xradia UltraXRM L200 (a, 5\,min exposure, Cu $\textrm{K}_\alpha$ radiation, by courtesy of Peter Kr\"uger, Fraunhofer IZFP Dresden), with the described transmission setup (b, 20\,min, 30\,kV tungsten spectrum), and the described reflection setup (c, 2\,min, 30\,kV tungsten). Details down to 100\,nm are clearly distinguishable in all three cases. d) and e) show phase contrast enhanced radiographs of low absorbing fibers (C and $\textrm{Al}_2\textrm{O}_3$ resp.) imaged with our transmission setup (upper) and at a synchrotron beamline (lower) and with the reflection setup. The effective propagation distance was 9.2\,mm and 4\,mm, resp. In terms of coherence these results are comparable.}
 \label{fig:Res}%
\end{figure}
To determine spatial resolution, a star pattern with structures down to 50\,nm was imaged with the presented setups and with a commercially available \Xray microscope, the Xradia UltraXRM L200, which is based on Fresnel zone plates (FZP). The resolution achieved with our microscopes is very comparable to the laboratory FZP microscope, however the contrast in our image is lower due to the fact, that we use a $30\Unit{kV}$ tungsten spectrum opposed to monochromatic Cu $\textrm{K}_\alpha$ radiation. The exposure time is longer for the geometric magnification in transmission mode because of the very high source-detector distance. By reducing the source-object distance in reflection mode a higher photon flux on the detector can be realized, thus reducing the exposure time significantly. In all three cases details down to 100\,nm are clearly distinguishable (see fig. \ref{fig:Res}a through c).\par
To demonstrate the phase contrast capabilities of the systems, low absorbing fibers were used. A 6\,\textmu m carbon fiber was imaged in transmission geometry and at a synchrotron beamline under comparable conditions (see fig.\ \ref{fig:Res}d). The edge enhancement due to phase contrast is comparable in both images. To test phase contrast imaging in the reflection setup, a 20\,\textmu m $\textrm{Al}_2\textrm{O}_3$ fiber was used. Even at high magnifications and thus low effective propagations distances edge enhancement is clearly visible.\par
\begin{figure}
	\centering
	\includegraphics[width=0.8\columnwidth]{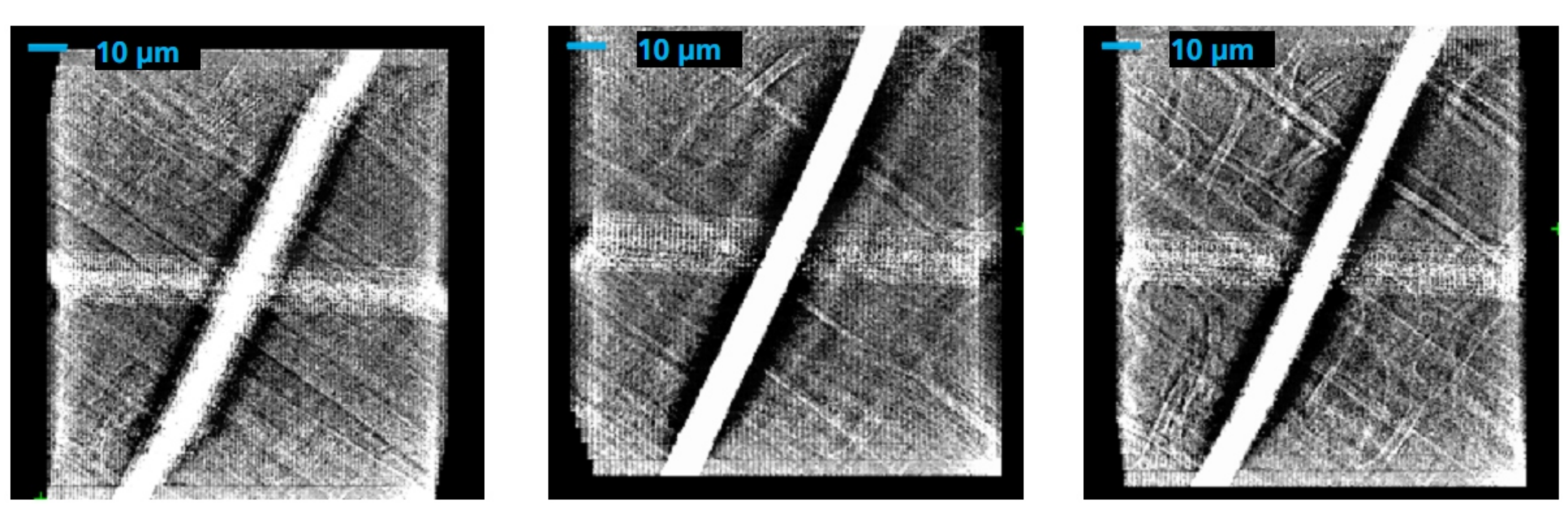}%
	\caption{Three slices of a volume reconstructed from five projections of a small piece of wood. At different levels, different features of the object are in focus (wire markers, cell walls). Alongside the visible wire photon starvation artifacts are visible. Exposure time was 5\,min per projection, voxel size is 500\,nm.}%
	\label{fig:LamiRes}%
\end{figure}

As a first test of the laminographic setup five projections of a small piece of wood marked with two metal wires on top were taken with linear detector/object displacements. A volume reconstruction was performed using an algebraic reconstruction technique. In fig.\ \ref{fig:LamiRes} three different reconstruction layers are shown. High resolution 3D imaging is possible with this setup without applying elaborate numerical treatment to the raw data.

\section{Summary and Outlook}
Both presented setups for \Xray microscopy based on geometric magnification show a spatial resolution of $100\Unit{nm}$ which is comparable to laboratory \Xray microscopes using synchrotron techniques. Even the very high resolutions achieved at synchrotron beamlines appear to be in reach. We also demonstrated high resolution laminography.\par
A detailed study of the resolution in 3D imaging is planned to also account for other influences on the quality of the reconstructed volume than the quality of the projection images, e.g.\ the precision of movement and stability.
 
\acknowledgments
The authors gratefully acknowledge the funding of the Bavarian State Ministry of Economic Affairs, Infrastructure, Transport and Technology which supports the project group "Nano-X-ray Systems for Material Characterization" and of the German Research Council (DFG) which, within the framework of its `Excellence Initiative` supports the Cluster of Excellence `Engineering of Advanced Materials` at the University of Erlangen-Nuremberg. Special thanks go to Dr.\ Peter Kr\"uger for providing the resolution test and images from the Xradia microscope.

\end{document}